\documentstyle[12pt,epsfig]{article}
\def\coskx{$cos$(k_xa_o)}
\def\cosky{$cos$(k_ya_o)}
\def\eplus{$e$^{i(k_x + k_y)a_o}}
\def\conjeplus{$e$^{-i(k_x + k_y)a_o}}
\def\eminus{$e$^{i(k_x - k_y)a_o}}
\def\conjeminus{$e$^{-i(k_x - k_y)a_o}}
\def\ethetax{$e$^{i\theta_x}}
\def\ethetay{$e$^{i\theta_y}}
\def\ethetaplus{$e$^{i\theta_{x+y}}}
\def\ethetaminus{$e$^{i\theta_{x-y}}}

\begin{document}
\def\bea{\begin{eqnarray}}
\def\eea{\end{eqnarray}}
\def\a{\alpha}
\def\b{\beta}
\def\d{\delta}
\def\l{\lambda}
\def\p{\partial} 
\def\#{\nonumber}
\def\r{\rho}
\def\la{\langle}
\def\ra{\rangle}
\def\e{\epsilon}
\def\n{\eta}
\def\hn{\hat{\eta}}
\def\break#1{\pagebreak \vspace*{#1}}
\def\byn{\frac{1}{n}}
\def\beq{\begin{equation}}
\def\eeq{\end{equation}}
\title{$Sr Cu_2 (BO_3)_2$: A Unique Mott Hubbard Insulator  $^{\dagger \; *}$  }
\author{B Sriram Shastry and Brijesh Kumar \\ Physics Department\\ Indian Institute of Science, Bangalore 560012,India   }
\maketitle
\date{18 March 2002}
\begin{abstract}
We discuss the recently discovered system $Sr Cu_2 (BO_3)_2$, a realization of an exactly solvable model 
proposed two decades earlier. We propose its interpretation as a Mott Hubbard insulator. The possible 
superconducting phase arising from doping is explored, and its nature as well as its importance for testing the RVB 
theory of superconductivity are discussed.
\end{abstract}

{\em Dedicated  to Professor Bill Sutherland on  occasion of his 60$^{th}$ birthday.  }
\section{Introduction}
Quantum spin systems are of great  current interest, as shown by this symposium, with roots in two distinct sources.
On the one hand the theory of model systems providing a rich variety of possibilities, 
 and on the other, the field of synthetic materials, which has generated
a vast number of systems, often close to theoretical models.
 As a result of this interplay, several interesting systems have been made in the laboratory,  
challenging our understanding by producing  not only the expected, but also on occasion, the unexpected.
Such a system that has caught attention recently is $Sr Cu_2 (BO_3)_2$,
a two dimensional $S=1/2$ isotropic Heisenberg antiferromagnet in two dimensions on  a particular lattice with 
the property that it is solvable exactly for the ground state. Indeed  it was solved two decades ago \cite{ss} by Sutherland 
and one of us. 
In this article we  summarize the story so far, and also explore possible interesting physics that could arise if 
this system  is doped. 

The situation of exactly solvable models in the area of statistical mechanics is rather limited.
There is a general feeling that the special models are non-generic and rare, and hence somewhat 
ornamental. Enlightened opinion\cite{mef}  has been more positive, and indeed
the role of some solvable  models is very well recognized. In contrast, the situation in condensed matter
physics  is very positive.  The interaction between new systems, new phenomena
and novel concepts has been rewarding. The table below gives a few examples of popular systems, their realizations
and the unique concepts associated  with them. 
\begin{center}
\begin{tabular}[2in]{p{.5in}|p{1.5in}|p{1.5in}|p{2in}}\hline
Year &{\bf Model Systems} & {\bf Realization} &{\bf New Concepts}\\ \hline
1930 &one dimensional Heisenberg Bethe AFM & CPC, CuO chains &Quantum Disorder, Spinons\\
1968 &one dimensional Majumdar Ghosh AFM&  (approximate mapping) $CuGe O_3$ & Broken Discrete Symmetries and Spinons\\
1969-87 &one dimensional $1/r^2$  Calogero, Sutherland, Haldane, Shastry systems & Parametric Correlations in Quantum Chaos& Spinons, Unusual Statistics\\
1969 & one dimensional Hubbard Model & Benzene,  Annulenes & Spin Charge Separation, Holons, Spinons,
SC Fluctuations from repulsion,  Mott Hubbard  Insulating state\\
1988& one dimensional Spin-1 Heisenberg AFM, Affleck, Kennedy, Lieb and Tasaki chain & Ni Chains, NENP & Haldane Spin Gap \\
 1990&  one dimensional n leg Heisenberg Ladders & Vanadates $Ca V_n O_{2n+1}$ & Integer vs non integer phenomena, Superconductivity from doping
 Insulators\\
1981 & two dimensional S=1/2 Shastry Sutherland model & $Sr Cu_2 (BO_3)_2$ & Dimer states, Magnetization Plateaus...\\ \hline
\end{tabular}  
\end{center}

The $1/r^2$ system in this table is different from the rest in that the physical realization comes from the world
of quantum chaos, the continuum model is the description of parametric correlations in chaotic systems. The sole two dimensional 
system in the list is the main concern of this article. It has  for long been unique in its very existence as a two dimensional member of
the family of solvable models. It is  particularly  surprising since it is a    
 model with  essential simplicity as evidenced by   the absence of crossed bonds.
It is now even more remarkable in that nature finds a way of fulfilling the conditions for solvability in the compound
  $Sr Cu_2 (BO_3)_2$. We discuss the origin of the model, its discovery in real life, some recent 
interesting developments in the physical properties, and some possible future directions.

\subsection{Origin of the model}
In view of the enormous current interest in the problem, and also questions from 
colleagues,  it may not be inappropriate to say 
a  few words on the Shastry Sutherland (SS) model on a special lattice, and how it came about. In 1980, I  (BSS) joined
the  University
of Utah  as a junior faculty member in the group
consisting of Professors D C  Mattis and B Sutherland.
 After an inspiring talk from Professor J R Schrieffer on polyacetylene, I mentioned to Professor Sutherland that
a clear  magnetic analog of polyacetylene ground state is the Majumdar Ghosh (MG) model, the one dimensional Heisenberg
 with a second neighbour interaction half as strong as the first. Professor Majumdar, my PhD advisor at TIFR
 in Bombay  earlier, had invented this model in an effort to go beyond the Bethe nearest neighbour
antiferromagnet ( AFM).
 The model was well known  to me, in spite of rather wise
 discouragement by Majumdar from working on Exactly Solvable models, as a discipline
unconnected with traditional topics in Solid State and Many Body Physics,
 in view of the almost  zero probability
of finding a new one!  I remember being surprised that   Sutherland, already then a {\it sensei} in the area 
of exactly solvable systems, had not come across this model! 
 In that characteristically  American way, there was no gap between learning
of a new thing and getting excited and plunging into. I also caught the
excitement  that I had carried, but so far resisted (!).   We first came up with
the soliton excitations of the MG model, the so called spinons, as isolated unpaired spin 1/2 propagating objects
in the midst of a sea of singlet dimerized spins. These were identical to the solitons of Schrieffer
in spirit, but fractionalized the spin degrees of freedom rather than charge. 
Such excitations have since  become  a paradigm in the post High Tc 
language of strongly correlated systems, where the fixed singlets of MG give way to dancing singlets, the Resonating
Valence Bond States envisioned by Anderson.   

\begin{figure}[h]
 \centerline{\epsfig{file=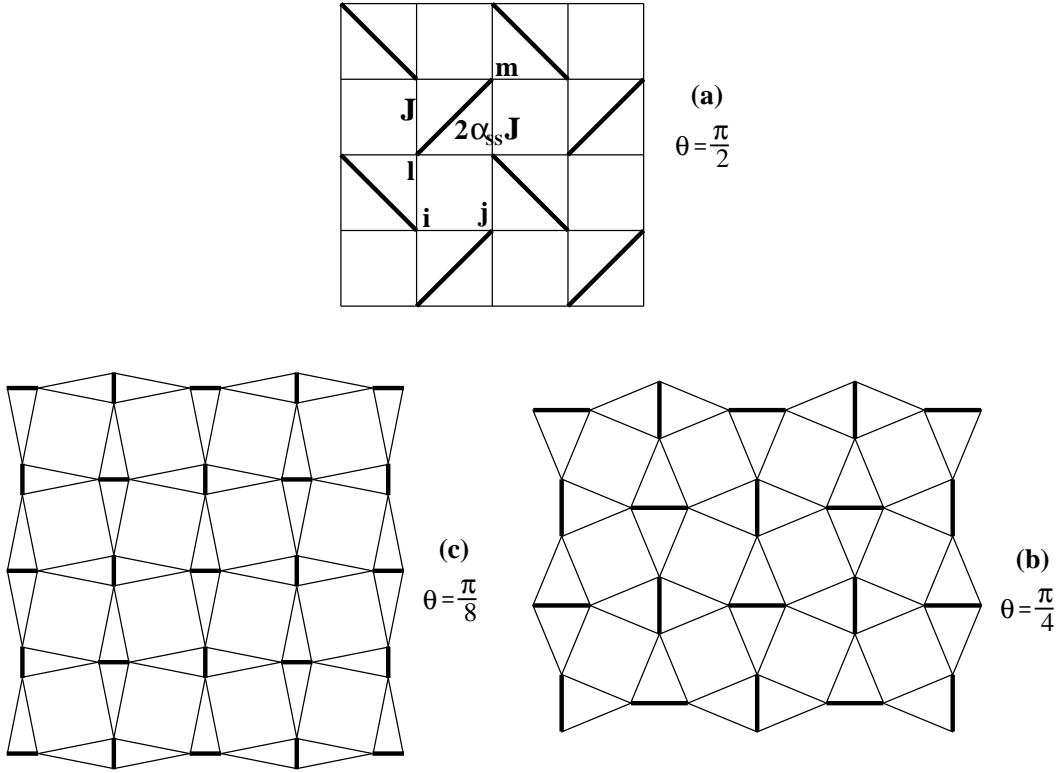,width=14cm}}
 \caption{The SS lattice. The angle $\theta$ is the apical angle for triangles that are the building blocks of the lattice, and by
continuously changing it, one generates different looking lattices with essentially identical topology. (a) represents the
original choice of SS, and (c) the case closest to the Copper lattice of $Sr Cu_2 ( BO_3)_2$.}
 \label{first}
\end{figure}

In an effort to go beyond one dimension to higher dimensions, we tried various things.  It was clear that a decomposition
into triangles was the key to the MG model, and there was no essential reason why this had to be only
one dimensional. The general point made was clear\cite{ss2}, the search for Integrable systems in higher dimensions is not very rewarding, the 
conditions for integrability seem  hard to satisfy in higher than one dimension, however, the search for exactly
solvable models ( for the ground state) is more promising {\it a priori}. In a $d$-dimensional Hilbert space there is a 
huge  number ( $\sim d^2/2$)of, 
 in general, non commuting operators that simultaneously share a given eigenstate, for example the dimer covering,
and the search boils down to states and operators that satisfy the somewhat subjective criterion of ``naturalness''.
 As a result
 we pondered for several weeks on likely systems
such as the two dimensional triangular lattice,  where it became clear that no dimer like
states work  since the triangles  share bonds with more than one other triangle.  One needed a lattice
where   for a given   triangle, no more than one bond is shared with another. 
This line of thinking led to the 
SS lattice shown in Fig.\ref{first}(a).

 The proof of the ground state is simple and worth repeating if only briefly.
The Hamiltonian can be written as a sum over triangles
\beq
H= J\Sigma_t H_t  =  J\left(\sum _{<i ,j>} \vec{S}_i.\vec{S}_j +   2 \alpha_{SS}  \sum _{<l , m>} \vec{S}_l.\vec{S}_m
\right)
\eeq
where the subscript $t$ refers to triangles, with $H_t= \alpha_{SS} \vec{S}_1.\vec{S}_2 + ( \vec{S}_1 + \vec{S}_2) . \vec{S}_3 $, ~ $\alpha_{SS}$ is the
bond strength parameter, sites $1,2$ refer to the two sites on the diagonal and $3$ the third site. Here and later 
we will denote the ``dimer'' bonds by $l,m$ and the non dimer nearest neighbours as $i,j$. 
The first, and remarkable point is that the dimer state $\psi = \prod_{l<m} [l,m]$  is an eigenstate of $H$.
Here the product runs over all dimers on the lattice, which must provide a covering of the lattice ( i.e. every lattice point
must occur once and only once in the product).
This happens because we can rearrange the operation of the Hamiltonian into two classes of terms, the wanted 
and the unanted terms. The wanted terms isolate the spin interactions on the dimer spins. Remarkably {\it all
unwanted terms have the form} $\vec{S}_j. ( \vec{S}_1 + \vec{S}_2)[1,2]$, for appropriate indices, which vanishes
on using the singlet property. By Rayleigh Ritz variational principle $E_{dimer} \geq E_0$, but by the Anderson
decomposition strategy we have a lower bound $E_0 \geq N_t e_t$. Happily 
the upper and lower bounds coincide for $\alpha_{SS} \geq 1$, and we are guaranteed that this is the ground state. Later 
work\cite{kk}  improves this lower bound on $\alpha_{SS}$ somewhat   to $ \sim 0.74$.

 These dimer ground states in two dimensions turn out to be surprisingly robust, for example the coupling constant $\alpha_{SS}$
is determined by an inequality rather than an equality, so there is an entire phase where the dimer states
are the ground state, unlike the one dimensional case where one has a solution at 
isolated points only.  This clearly 
greatly increases the probability of finding such models realized in nature, whereas in one dimension we should
only expect proximities. Further the ground state
is insensitive to the spin space isotropy of the underlying Hamiltonian, and one has the strange situation where the ground state
has a greater symmetry than of the Hamiltonian!  The  phrase ``superstability'' \cite{ss2}  describes this kind of
robustness shared by most of the dimer ground state systems. An example   of robustness comes from later in the story,
where we find that the ground state of stacks of the SS lattice, rotated by $\pi/2$ and coupled by vertical
spin interactions, a model that describes the real  3-d material $Sr Cu_2(BO_3)_2$, ``magically'' manages to have 
the dimer state as the true ground state\cite{mu}!

\subsection{The system $SrCu_2(BO_3)_2$} Almost two decades later,  Kageyama and coworkers at the ISSP in Tokyo found that $SrCu_2 (BO_3)_2$, synthesized  earlier 
in 1991  by Smith and Keszler, had  very unusual properties. The spin $1/2$ moments of Copper living in well isolated two dimensional layers seemed to lock up into
singlets, and a clear spin gap behaviour was observed by NMR.  They concluded that this is a unique system, the first
truly two dimensional spin gapped system with $S=1/2$. The data was analyzed by  Mihayara and  Ueda (MU)\cite{mu}, 
who realized that
 the physical system was describable by  an exactly solvable model.  They proposed the model, found its solution, and 
then realized that it was essentially (topologically) the same model as SS, but looked different due to the details of the
lattice. The Copper lattice is shown in Fig.\ref{first}(c), and by opening up the angle $\theta$ continuously, one reaches 
the SS lattice ( with $\theta= \pi/2$ ).
 An intermediate value of $\theta= \pi/4$ in Fig.\ref{first}(b)  aids the imagination. Changing the
angle $\theta$ clearly preserves the orthogonality of the ``dimer'' bonds but changes the bond length relative to the 
inter dimer bond lengths. This is  crucial, since the criterion for solvability $\alpha_{SS} \geq\sim 0.74 $ becomes
realizable only in this picture. A nice visualization of this deformation  is available courtesy Dr H Kageyama at 
http://www.issp.u-tokyo.ac.jp/labs/mdcl/ueda/kage/head.html.
 The Hamiltonian is written by MU  and some recent papers
as $H=J[  \alpha_{MU}  \sum _{i<j} \vec{S}_i.\vec{S}_j +      \sum _{l < m} \vec{S}_l.\vec{S}_m] $, with
$J' \equiv J  \alpha_{MU}$,
 and hence we clearly have $\alpha_{SS}\cdot \alpha_{MU}= \frac{1}{2}$. The current estimates of ($J, \; \alpha_{MU}$)
using experimental data on $SrCu_2(BO_3)_2$  range from \cite{mu}  ($ 85^0K , \; 0.635$)
to \cite{knetter} ($ 71.5 ^0K , \; 0.603$).
 Thus $\alpha_{SS} \sim 0.78$ is 
perilously close to, but on the safe side of the phase boundary at $\sim 0.74$.

\subsection{Recent Developments} We next mention a few of the very large number of  papers that have
been written recently, with apologies in advance for possible incompleteness.
After the discovery of the spin gap, neutron scattering 
has confirmed the absence of magnetic LRO and inelastic scattering has
given clear indication of a flat dispersionless triplet excitation mode at about 3 meV, as well as
of  many branches of dispersing bound states of triplet excitations\cite{neutron}.
 NMR experiments were the first to show the spin gap\cite{nmr} at about 30 K.
 ESR experiments show the presence of  a second gap at about 4.7 meV which implies  a substantial  binding
energy of two triplets\cite{esr}. Raman studies show a singlet bound state at about  3.7 meV \cite{raman}. 
The  magnetic exchange constants, as mentioned are in the  
60-80 K range. This is convenient for exploring with available pulsed high magnetic field experiments, which
reveal\cite{nmr, plateauexp} the surprising existence of magnetic plateaus at $M/M_s= 1/4, 1/3, 1/8..$.  There is
interesting data on the effect of magnetic excitations on phononic thermal conductivity\cite{thermal}. 

Thus a large
set of experiments have already been done, and provide many constraints on  the theory. We should
mention that the knowledge of the  exact solution of the ground state does not give much 
insight into the excitations in this class of systems. One knows that in general terms, the singlet dimers
can be broken into triplets, and that isolated triplets find it hard to propagate on the SS lattice due to
its topology. This leads to flat bands of triplet excitations, i.e. very massive objects,  consistent with neutron data.
 Pairs of triplets, however, escape the topological constraints, much as holes in the N\`{e}el Antiferromagnet, and move about quite freely. Thus one has
kinetic binding and the bound state has substantial dispersion\cite{excitations}.

In a sense the unexpected and new physics so far has been the presence of these plateaus\cite{plateauexp}. These are unique
in that they are the first two dimensional plateaus seen, and have attracted considerable interest.
Several possible scenarios have been suggested to explain these.
One picture is that of massive
triplet excitations  acting as hardcore bosons, that hop as well as interact. The effective interactions are strong due to
large mass, and Wigner crystallization is proposed to explain the plateaus\cite{plateaus1}.
 Alternatively one can view this
plateau formation as the Quantum Hall Effect of  hard core bosons, and a Chern Simons type field theory provides a fair
description. The structure of the Hofstadter spectrum on the SS lattice is reflected in the plateaus\cite{plateaus2}.
At the moment it is not easy to reach a conclusion as to the best interpretation, especially since the experiments 
do not show plateaus that have anything like the precision of the Quantum Hall Effect.

\begin{figure}[h]
 \centerline{\epsfig{file=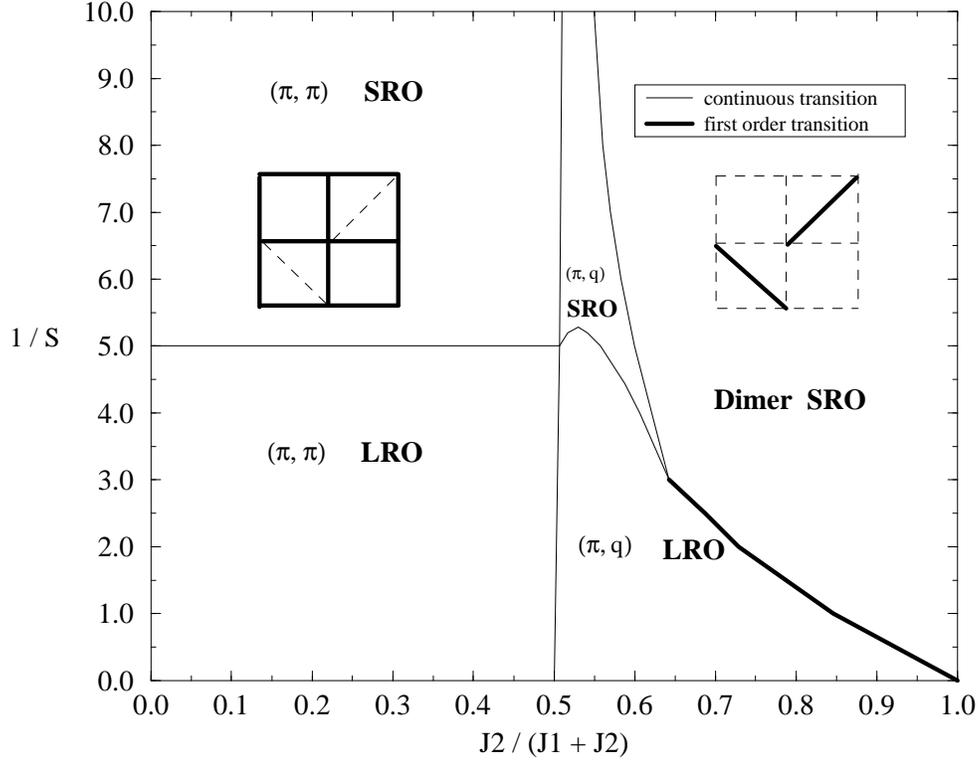,width=12cm}}
 \caption{The phase diagram of Ref. \cite{cms} in the large N limit. The abscissa may be read as 
          $2 \alpha_{SS}/( 1 +2 \alpha_{SS})$.}
 \label{second}
\end{figure}

The phase diagram at zero field has come in for close scrutiny by several authors, using series expansion ideas\cite{kk,oitmaa},
rigorous bounds\cite{uwe},  large N field theories\cite{cms}, as well as effective field theories 
of  bosonic dimers\cite{carpentier}.  Exotic and unusual intermediate phases are suggested by these
studies, including\cite{cms} a ``topologically ordered phase with deconfined $S=1/2$ spinons, which should give 
rise to an exotic superconductor with anomalous flux properties under doping''.
 An early paper by Albrecht and Mila\cite{am} discusses the transition
between the dimer and AFM phases using Schwinger Bosons, and concludes that it should be first order. 
 A  qualitatively new intermediate spin liquid phase with power law correlations 
has been proposed by Koga and Kawakami\cite{kk}. The phase diagram in 
Fig.\ref{second} is from reference \cite{cms} for a large $N$ theory, and  represents a possible set of phases with various
 kinds of magnetic order. The Koga-Kawakami phase may be viewed as the SRO $(\pi,q)$ phase.

We should also mention new theoretical models that are generalizations of the SS ideas to
higher dimensions and other systems\cite{mh,theoretical}. The  beautiful
model of M\"{u}ller-Hartmann, Singh, Knetter and Uhrig
\cite{mh} has a new set of exchange interactions added to the SS model, and had a very rich set of 
constants of motion.   It is very tractable, giving rise to magnetization plateaus that are similar to but not identical
to the experimental ones.

\section{ Mean Field Theory of   the Doped Dimer  Superconductor, a Possible  test  of RVB}

In this section we discuss the possibility of doping the dimer state,  and what one may
expect from it.  Firstly we remark that the insulating  dimerized ground state of  $SrCu_2 ( BO_3)_2$  
may usefully be considered as a Mott Insulating state of an underlying Hubbard model.  To see this  consider
the Hubbard model on the SS lattice with 
\beq
H= - t \sum_{<i j>,\sigma} (c^\dagger_{i \sigma} c_{j\sigma}+ h.c.) - \alpha t \sum_{<l,m>,\sigma} ( c^\dagger_{l,\sigma} c_{m,\sigma} + h.c. )
+ U \sum_{r} n_{r, \uparrow} n_{ r, \downarrow}
\eeq
where the (somewhat overused) symbol
$\alpha$ represents the ratio of hoppings on the two kinds of bonds, and the Hubbard interaction term
sums over all types of sites. Clearly the superexchange argument fixes it in terms 
of the ratio of the exchange parameters via $\alpha^2= 2 \alpha_{SS}$, and we note that $\alpha =\pm 1.25$ using the
insulating state estimates, with the sign undetermined.
In the non-interacting limit  the band structure is interesting, we have four subbands, with the extrema of
two of them touching
quadratically at the zone center. At half filling, one has
four electrons per unit cell and the system is a semi-metal  with a finite density of states, and thus it has typical metallic behaviour such as 
a linear specific heat.  A parallel may be drawn with the semi-metallicity of graphite on the hexagonal lattice
with two electrons per unit cell and 
also of a fiduciary $MgB_2$ with well separated planes.  In the case of graphite however, one has a ``Dirac like''  linear spectrum, and hence  the density of states near the 
``fermi point'' , i.e the contact point  vanishes.   

This semi-metal  becomes an insulator at large enough $U$, undergoing a transition to the dimerized  state that does not break rotation invariance nor the lattice translation symmetry, and may be called a Mott transition
in the same sense as that of the one dimensional Hubbard model at half filling at infinitesimal $U$. 
Since the large $U$ behaviour is exactly known, namely the dimer ground state, further terms in the $t/U$
expansion beyond superexchange should be useful in elucidating the nature of the Mott transition here \cite{nozieres}.
This transition has not yet been studied in literature. Starting from the semi-metal and turning on $U$,
one may either have a level crossing transition to the gapped insulator, or more interestingly a continuous opening 
of the charge gap. In the gapped insulating phase, the four spin correlation function pertaining to dimer order
 $<S_a S_b S_c S_d> \sim <S_a S_b> < S_c S_d>$, and thus there is ODLRO in this correlation function without any obvious symmetry that is broken.
With this distinction, without necessarily a major difference, we may refer
to $SrCu_2 ( BO_3)_2$ as a Mott insulator. 

Having this realization of the Mott  Hubbard insulator, we naturally enquire if the philosophy of the RVB theory of Superconductivity
due to Anderson applies here. This theory is built upon the idea that repulsive interactions of the Hubbard type
lead to superconductivity via the intermediate step of superexchange, or Heisenberg interactions in the insulating state.
The superexchange leads to singlet pairing between electrons of opposite spin, and these pairs are analogous to 
the Cooper pairs, but are localized due to the Mott-Hubbard gap. Under doping the Mott-Hubbard gap collapses, these preexisting pairs 
propagate freely, and lead to superconductivity. In the present case, the Cooper pairs at half filling should be
viewed as the dimer-singlets, which on doping should move around by the same logic, and lead to superconductivity.
Since the values of exchange are smaller by an order of magnitude from those in High Tc systems, we expect
lower Tc's, say tens of degrees K, but accompanied by the characteristic signature of singlet pairing and also of definite
phase relations of Cooper pairs on the bonds, analogous to the d-wave pairing. While this theory is 
remarkably effective in providing a comprehensive view point, it still lacks unambiguous experimental support or
 a rigorous mathematical foundation, and one
would  welcome  other supports to its validity or otherwise. In this context we work out in this section the
mean field theory of a fiducial doped $Sr Cu_2 ( BO)_3$, and calculate some characteristics of the
proposed superconducting compound.

Before doing so, let us note that doping can be of either chemical type, as in say $Sr_{1-x} M_x Cu_2 ( BO_3)_2$ with
a monovalent alkali  $M$  or a trivalent lanthanide. However, one interesting possibility is suggested by the 
comparison of $MgB_2$ with graphite. One learns that $MgB_2$ is isoelectronic with graphite, but avoids being 
a semimetal by dispersing the bands in the direction transverse to the two dimensional sheets, it self-dopes by
decreasing the transverse lattice constant. It is possible that a    divalent element like $Mg$ in place of $Sr$
 with a smaller ionic  radius
could decrease the transverse lattice constant of $Sr Cu_2 ( BO_3)_2$ sufficiently so 
that it would have substantial transverse dispersion. We should  clarify that unlike $Mg B_2$ which appears to be
a case of phonon mediated superconductivity \cite{mgb2}, we are examining the case for a non phononic
mechanism, the RVB mechanism for doped   $Sr Cu_2 ( BO_3)_2$. Gate charging might be another attractive 
possibility. We now turn to the calculation proper.

\subsection{RVB type mean field theory on the SS lattice}
\begin{figure}[h]
 \centerline{\epsfig{file=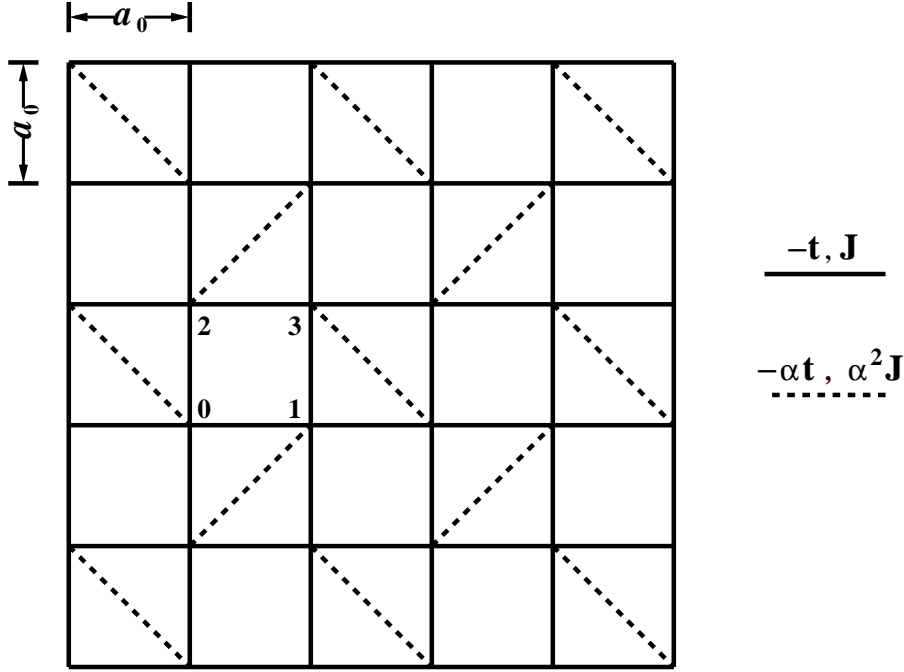,width=12cm}}
 \caption{The Shastry-Sutherland lattice. Also shown is the labeling of sites in a unit cell of SS lattice, as
          used in the text.}
 \label{SS_lattice}
\end{figure}

We next present the mean field theory of a $t-J$ type model on the SS lattice.
The hopping amplitudes and the exchange integrals on SS lattice are as shown in Fig.(\ref{SS_lattice}). 
The nearest neighbour (n.n.) hopping amplitude is $-t$, and the next nearest neighbour (n.n.n.) hopping amplitude
 is $-\alpha t$ where $\alpha$ is a dimensionless number. The exchange couplings, $J$ and $J^\prime$ along 
the n.n. and n.n.n. directions respectively, are such that $J^\prime=\alpha^2 J$, as governed by the large $U$ 
physics of the Hubbard model on SS lattice. Thus $\alpha^2 = 2\alpha_{SS}$ of the previous section.
The $t-J$ type model, thus arrived at is an appropriate generalization of the original SS
model, in order to deal with doping. In the following, we will first describe the tight binding band structure
 of the free electrons on the SS lattice. Then, we will do the mean field theory of the interacting model 
and discuss its implications for superconductivity, in a manner analogous to the early RVB mean field theories of
$t-J$ model on a square lattice done in the context of high-T$_C$ superconductivity\cite{BZA, kotliar_liu}. 

\paragraph{The band-structure :}The SS lattice has a periodicity of $2a_o$, both along $\hat{\mbox{x}}$ as well as 
$\hat{\mbox{y}}$ directions, where $a_o$ is the lattice constant. With each site contributing just one relevant
 orbital, the tight-binding model on SS lattice is described by a four band hamiltonian given below.
\begin{equation}
H_{t} = \sum_{{\bf k},\sigma}
\left[ c_0^\dagger({\bf k})\;\;c_1^\dagger({\bf k})\;\;c_2^\dagger({\bf k})\;\;c_3^\dagger({\bf k})
\right]_\sigma {\bf T(k)}
\left[\begin{array}{c} c_0({\bf k}) \\ c_1({\bf k}) \\ c_2({\bf k}) \\ c_3({\bf k}) \end{array}\right]_\sigma
\label{Ht}
\end{equation}
Here, $\sigma = \uparrow$ or $\downarrow$, and the wave-vector, ${\bf k} = (k_x, k_y)$, is such that 
$\frac{-\pi}{2a_o} \le k_x, k_y \le \frac{\pi}{2a_o}$. The subscripts, $0, 1, 2, 3$, refer to four different site
 within a unit cell. The dispersion matrix, ${\bf T(k)}$, is a $4\times 4$ hermitian matrix as given below.
\begin{equation}
 {\bf T(k)} = -t\left[\begin{array}{cccc}
                       0 & 2\;\coskx & 2\;\cosky & \alpha\;\eminus\\
                       2\;\coskx & 0 & \alpha\;\eplus & 2\;\cosky\\
                       2\;\cosky & \alpha\;\conjeplus & 0 & 2\;\coskx \\
                       \alpha\;\conjeminus & 2\;\cosky & 2\;\coskx & 0
                      \end{array}
               \right]
\label{T_matrix}
\end{equation}

\begin{figure}[h]
 \centerline{\epsfig{file=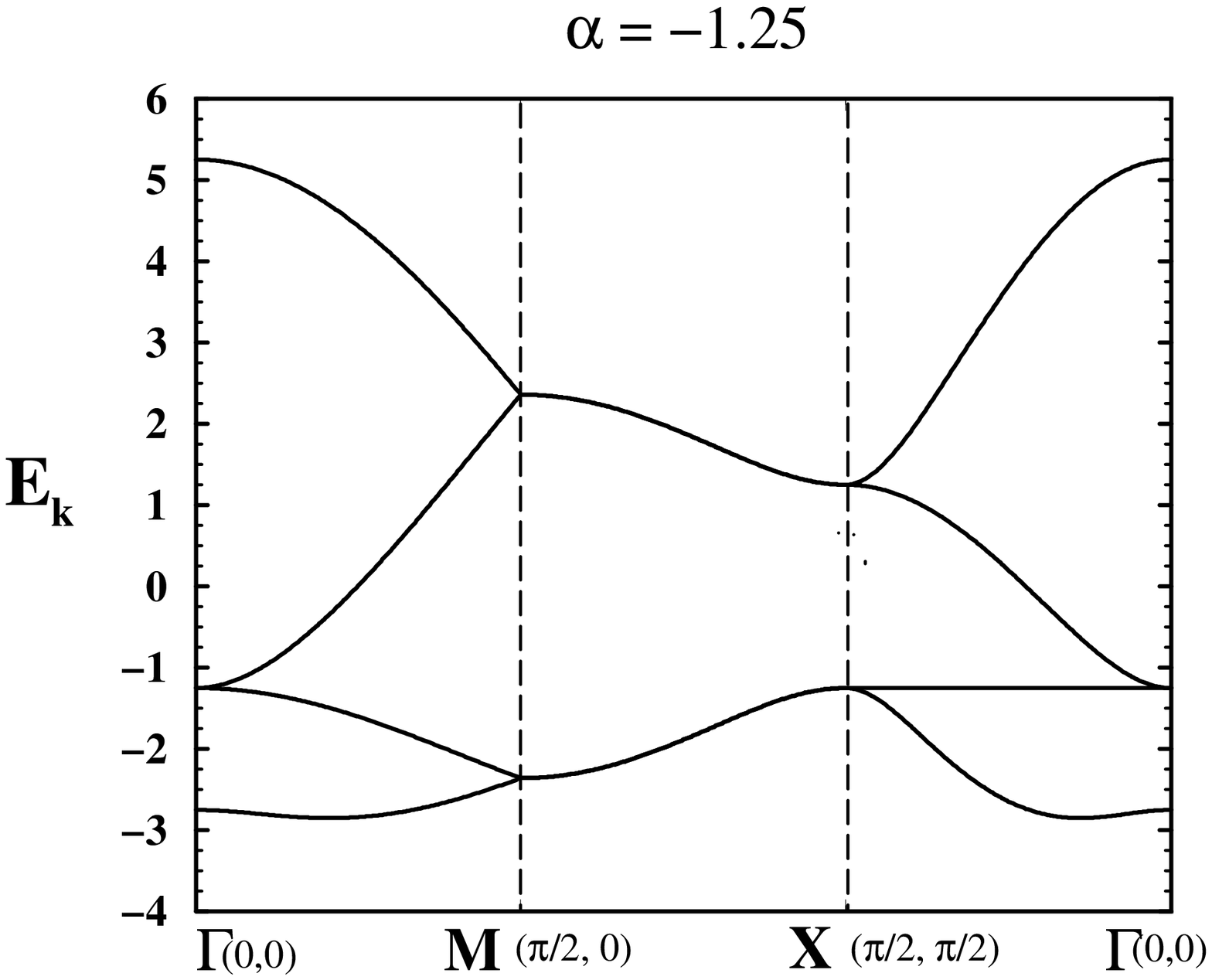,width=8.5cm}\hfill 
             \epsfig{file=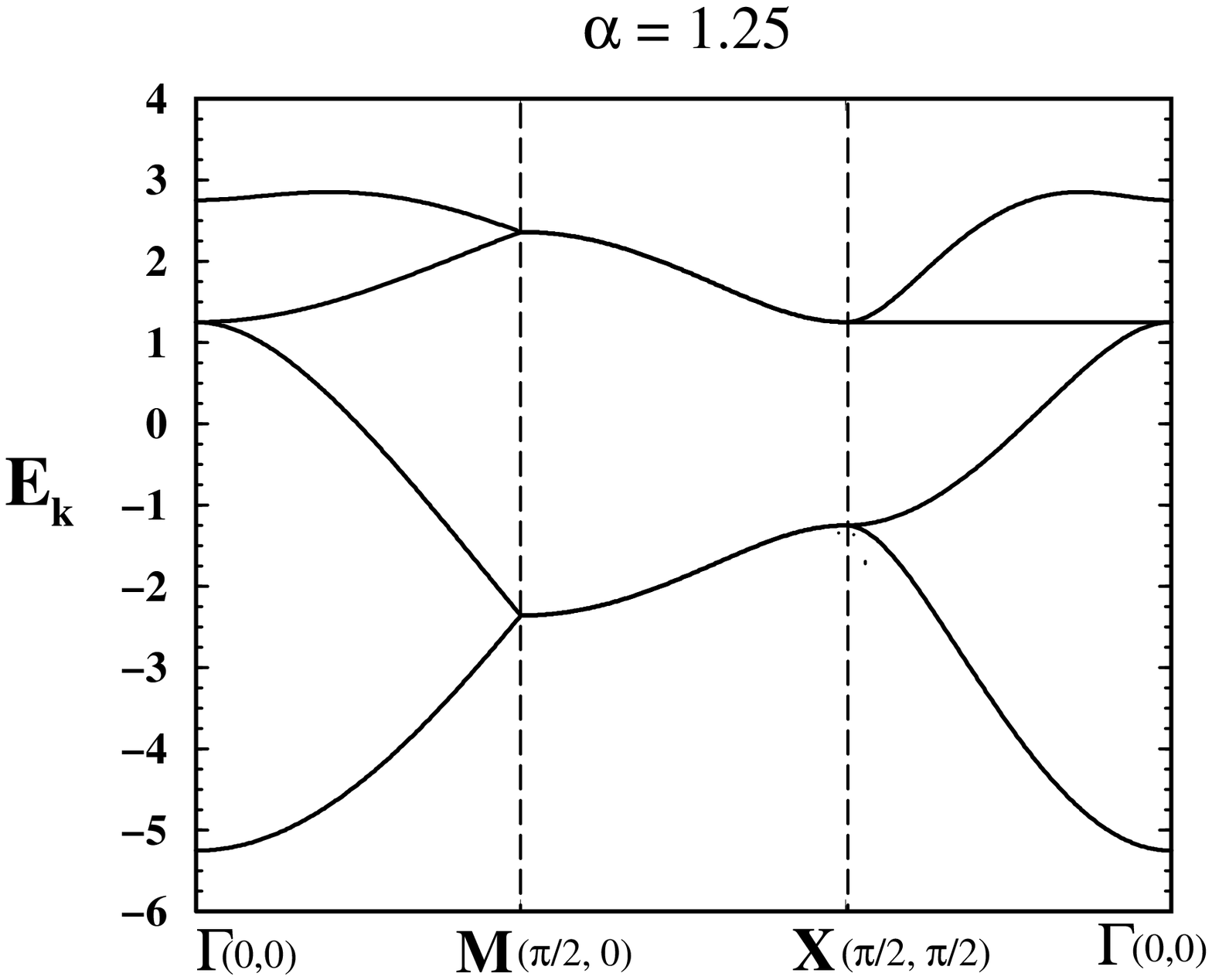,width=8.5cm}}
 \caption{The band-structure of free electrons on the SS lattice. The wave vectors are written in units of
          $\frac{1}{a_o}$. Notice that the band-structure is odd with respect to $\alpha$.}
 \label{band}
\end{figure}

The band-structure for $|\alpha| = 1.25$ is shown in Fig.(\ref{band}). This value of $\alpha$ is taken from
the studies on SrCu$_2$(BO$_3$)$_2$, where the values of $J$ and $J^\prime$ are extracted by fitting the
 experimental data with the orthogonal dimer model. What is known from the 
experiments is $\alpha^2$, and not $\alpha$. This leaves us with the ambiguity of sign of $\alpha$,
and hence we have considered both positive as well negative values of $\alpha$.

Let us make a few essential observations regarding the band-structure. Firstly, the system is a semi-metal at
half filling, since the middle two bands touch each other at the zone centre. Secondly, 
there is a band which is flat along the X$\Gamma$ symmetry direction in the Brillouin Zone.
This band gives rise to a severe van Hove singularity at $\alpha$. 
Thirdly, the values of band energies at zone centre are $(-4-\alpha)$, $\alpha$, $\alpha$ and $(4-\alpha)$. For 
 $|\alpha| > 2$, the middle two bands no more touch each other, and there is a finite band gap which makes it a
 band insulator at half filling. Since $\alpha$ for the material of real interest is roughly 1.25, we have not 
tried to discuss other values of $\alpha$.

\begin{figure}
 \centerline{\epsfig{file=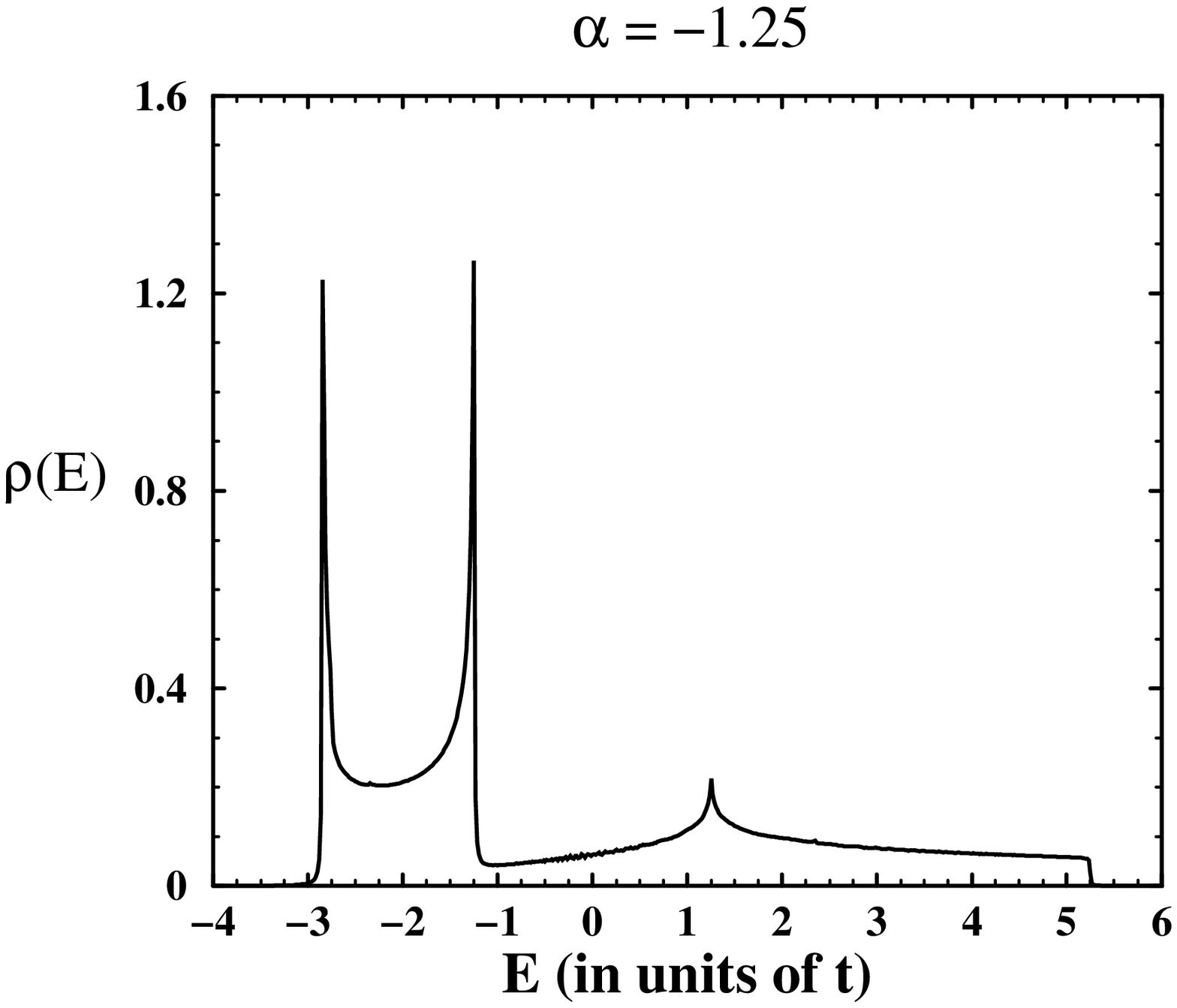,width=8.5cm}\hfill 
             \epsfig{file=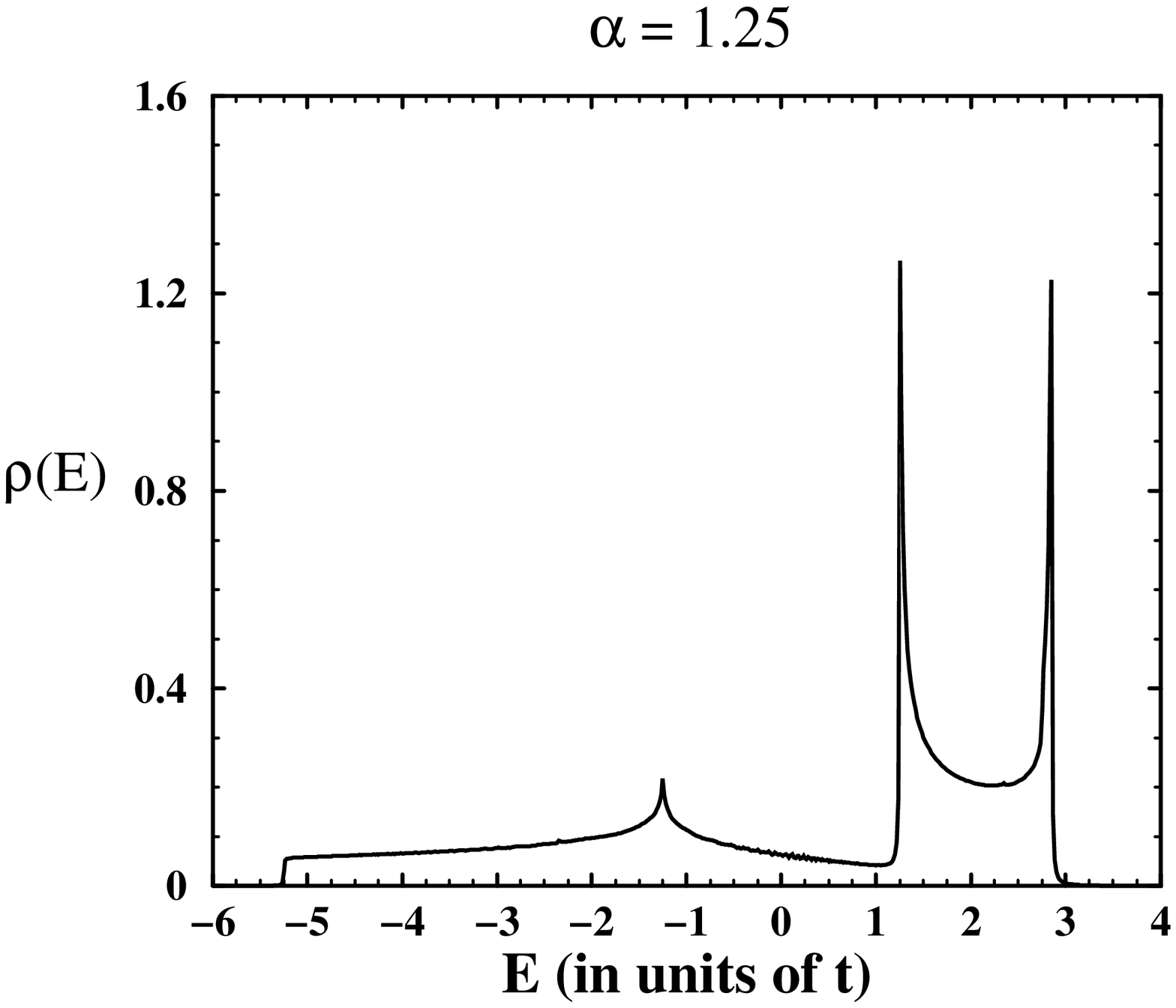,width=8.5cm}}
 \caption{The density of single-particle electronic states on Shastry-Sutherland lattice.}
 \label{dos}
\end{figure}

Fig.(\ref{dos}) shows the non-interacting single particle density of states on SS lattice for both negative as 
well positive values of $\alpha$. When we hole-dope the system to take it away from half filling, it is expected 
to behave differently for positive and negative $\alpha$, since the flat band influences the case only when 
 $\alpha$ is negative.

\paragraph{The mean field hamiltonian :} The $t-J$ hamiltonian on SS lattice can be written as:
\begin{equation}
 \tilde{H} = {\cal P}H_t{\cal P} + H_J - 
              t\mu\sum_{{\bf k},\sigma}\sum_{p=0}^{3} c_{p,\sigma}^\dagger({\bf k})c_{p,\sigma}({\bf k})
 \label{H_tilde}
\end{equation}
The first term in $\tilde{H}$ accounts for the projected hopping. It is essentially $H_t$ as given in
equation(\ref{Ht}), but with projection operator ${\cal P}$ which suppresses the double occupancy of any site
 (due to large Hubbard $U$). At a simple level, the effect of ${\cal P}$ can be brought in by replacing $t$ by 
$\delta t$.  Here, $\delta (= 1-n)$, is the number of holes per site, and $n$ is the electron filling per
site. The last term is the chemical potential, $\mu t$, times the total number of electrons. 
Here, $p$ is the site (or the orbital) label within a unit cell. The second term in equation(\ref{H_tilde}),
 $H_J$, which accounts for the interaction among electrons can be written as:
\begin{equation}
 H_J = J\left\{\sum_{n.n.} + \;\; \alpha^2\sum_{n.n.n.}^\prime \right\}\left({\bf S(r)}\cdot{\bf S(r^\prime)} - 
             \frac{\hat{\mbox{n}}({\bf r})\hat{\mbox{n}}({\bf r}^\prime)}{4}\right)
\label{HJ}
\end{equation}
Here, ${\bf r}$ and ${\bf r^\prime}$ are the site labels, and $\hat{\mbox{n}}(\bf r)$ denotes the number operator
at site ${\bf r}$. The summation is pairwise in ${\bf r}$, ${\bf r^\prime}$. The primed summation denotes
the sum of only those pairs of n.n.n. sites which are allowed by the connectivity of the SS lattice.
The operator, $\left({\bf S(r)}\cdot{\bf S(r^\prime)} -
\hat{\mbox{n}}({\bf r})\hat{\mbox{n}}({\bf r}^\prime)/4\right)$, can also be written as $-\frac{1}{2}{\bf
b^\dagger (r,r^\prime)}{\bf b(r,r^\prime)}$, which provides the basis for mean field decoupling of
$H_J$ in the off-diagonal channel. The operator, ${\bf
b(r,r^\prime)}=c_{\downarrow}({\bf r})c_{\uparrow}({\bf r}^\prime) - c_{\uparrow}({\bf r})c_{\downarrow}({\bf
r}^\prime)$, is the singlet bond operator. 

Let us define an off-diagonal or the pairing mean field, $\left<{\bf b(r,r^\prime)}\right>$, in the following
way.
\begin{equation}
 \left<{\bf b(r,r^\prime)}\right> = \left\{ 
       \begin{array}{ll}
        \Delta\ethetax & \;\;\;\;\mbox{for}\;\; {\bf r-r^\prime} = \pm a\hat{\mbox{x}}\\
        \Delta\ethetay & \;\;\;\;\mbox{for}\;\; {\bf r-r^\prime} = \pm a\hat{\mbox{y}}\\
        \Delta^\prime\ethetaplus & \;\;\;\;\mbox{for}\;\; {\bf r-r^\prime} = \pm a(\hat{\mbox{x}}+\hat{\mbox{y}})\\
        \Delta^\prime\ethetaminus & \;\;\;\;\mbox{for}\;\; {\bf r-r^\prime} = \pm a(\hat{\mbox{x}}-\hat{\mbox{y}})
       \end{array}
                                    \right.
\label{meanfield}
\end{equation}
The phases, $\theta_x$, $\theta_y$, $\theta_{x+y}$ and $\theta_{x-y}$, as well as the amplitudes, $\Delta$ and
$\Delta^\prime$, are all independent of the coordinates. Hence, we are considering a uniform case. With this
choice of the order parameter, we decouple $H_J$. The corresponding mean field hamiltonian can be written as:
\begin{equation}
 \tilde{H}^{MF} = \tilde{H}_t + H_J^{MF} + L(4J\Delta^2 + J^{\prime}{\Delta^{\prime}}^2)
\end{equation}
where $L$ is the number of unit cells. In order to write $\tilde{H}_t$ and $H_J^{MF}$ conveniently, we 
introduce a notation. Let us define the Nambu operators, $\Psi_{C\uparrow}({\bf k})$ and
 $\Psi_{R\downarrow}({\bf -k})$ in the following way.
\begin{eqnarray}
 \Psi_{C\uparrow}({\bf k}) & = & \left[\begin{array}{c} 
                                        c_{0\uparrow}({\bf k}) \\ 
                                        c_{1\uparrow}({\bf k}) \\ 
                                        c_{2\uparrow}({\bf k}) \\ 
                                        c_{3\uparrow}({\bf k})
                                       \end{array}
                                \right] \\
&& \nonumber \\
 \Psi_{R\downarrow}(-{\bf k}) & = & \left[c_{0\downarrow}(-{\bf k})\;\;
                                          c_{1\downarrow}(-{\bf k})\;\;
                                          c_{2\downarrow}(-{\bf k})\;\;
                                          c_{3\downarrow}(-{\bf k})
                                   \right]
\end{eqnarray}
The subscripts, $C$ and $R$, indicate that $\Psi_{C\uparrow}({\bf k})$ is a column vector and
$\Psi_{R\downarrow}({\bf -k})$ is a row vector. In this notation, $\tilde{H}_t$ can be written as :
\begin{equation}
 \tilde{H}_t = \sum_{\bf k}\left\{ \mbox{tr}\left\{ {\bf \tilde{T}(-k)}\right\} + 
 \left[ \Psi^\dagger_{C\uparrow}({\bf k})\;\; \Psi_{R\downarrow}(-{\bf k}) \right]
 \left[ \begin{array}{cc} 
         {\bf \tilde{T}(k)} & {\bf 0}\\ 
         {\bf 0} & -{\bf \tilde{T}(k)}
        \end{array} 
\right]
 \left[ \begin{array}{c} 
         \Psi_{C\uparrow}({\bf k})\\ 
         \Psi^\dagger_{R\downarrow}(-{\bf k})
        \end{array}
\right]
\right\}
\label{phtp}
\end{equation}
Here, ${\bf \tilde{T}(k)}$ is essentially same as the dispersion matrix, ${\bf T(k)}$, except that the chemical 
potential forms its diagonal elements, and all the off-diagonal entries have a factor of hole doping, $\delta$, 
in order to account for the projection.
\begin{equation}
 {\bf \tilde{T}(k)} = -t\left[\begin{array}{llll}
                                \mu & 2\delta\;\coskx & 2\delta\;\cosky & \delta\alpha\;\eminus\\
                                2\delta\;\coskx & \mu & \delta\alpha\;\eplus & 2\delta\;\cosky\\
                                2\delta\;\cosky & \delta\alpha\;\conjeplus & \mu & 2\delta\;\coskx \\
                                \delta\alpha\;\conjeminus & 2\delta\;\cosky & 2\delta\;\coskx & \mu
                      \end{array}
               \right]
\label{Ttilde}
\end{equation}
With the same notation, $H_J^{MF}$ can be written as :
\begin{equation}
 H_J^{MF} = \left[\Psi_{C\uparrow}^\dagger({\bf k})\;\; \Psi_{R\downarrow}(-{\bf k})\right]
                  \left[\begin{array}{cc}
                         {\bf 0} & {\bf D(k)}\\
                         {\bf D^\dagger(k)} & {\bf 0}
                        \end{array}
                  \right]
                  \left[\begin{array}{c}
                         \Psi_{C\uparrow}({\bf k}) \\
                         \Psi_{R\downarrow}^\dagger(-{\bf k})
                        \end{array}
                  \right]
\label{HJMF}
\end{equation}
where ${\bf D(k)}$ is a non-hermitian $4\times 4$ matrix as given below.
\begin{equation}
{\bf D(k)} = -\left[\begin{array}{llll}
                    0 & J\Delta\ethetax\coskx & J\Delta\ethetay\cosky &
                        \frac{J^\prime\Delta^\prime}{2}\ethetaminus\eminus \\
                    J\Delta\ethetax\coskx & 0 & \frac{J^\prime\Delta^\prime}{2}\ethetaplus\eplus &
                        J\Delta\ethetay\cosky \\
                    J\Delta\ethetay\cosky & \frac{J^\prime\Delta^\prime}{2}\ethetaplus\conjeplus & 0 &
                        J\Delta\ethetax\coskx \\
                    \frac{J^\prime\Delta^\prime}{2}\ethetaminus\conjeminus & J\Delta\ethetay\cosky &
                        J\Delta\ethetax\coskx & 0 
                   \end{array}
            \right]
\label{Dmatrix}
\end{equation}
Here, $J^\prime = \alpha^2 J$ as mentioned earlier. Finally, we write the $\tilde{H}^{MF}$ as :
\begin{eqnarray}
\tilde{H}^{MF} & = & \sum_{\bf k} \left[\Psi_{C\uparrow}^\dagger({\bf k})\;\; \Psi_{R\downarrow}(-{\bf k})\right]
               \left[\begin{array}{cc}
                      {\bf \tilde{T}(k)} & {\bf D(k)}\\
                      {\bf D^\dagger(k)} & -{\bf \tilde{T}(k)}
                     \end{array}
              \right]
               \left[\begin{array}{c}
                      \Psi_{C\uparrow}({\bf k}) \\
                      \Psi_{R\downarrow}^\dagger(-{\bf k})
                     \end{array}
              \right] \nonumber \\
        &   & + L\left(4J\Delta^2 + J^\prime{\Delta^\prime}^2 - 4t\mu\right)
\label{HtildeMF}
\end{eqnarray}
Let us denote the matrix \(\left[ \begin{array}{cc} {\bf \tilde{T}(k)} & {\bf D(k)}\\
{\bf D^\dagger(k)} & -{\bf \tilde{T}(k)} \end{array}\right] \) by {\bf A(k)}. 
It is an $8\times 8$ symplectic, hermitian matrix whose 
eigenvalues are real and occur in pairs. That is, an eigenvalue's negative is also an eigenvalue.
\paragraph{The mean field free energy and the self-consistent equations :}
The grand canonical free energy, $\Phi$, at a given temperature $T$, for the mean field hamiltonian described
above is,
\begin{equation}
 \Phi = 4L\left(J\Delta^2 + \frac{J^\prime{\Delta^\prime}^2}{4} - t\mu\right) - 
        \sum_{\bf k}\sum_{j=1}^4\left\{E_j^+({\bf k}) + 
                                       \frac{2}{\beta}\mbox{Log}\left(1+\mbox{e}^{-\beta E_j^+({\bf k})}\right)
                                 \right\}
\label{free}
\end{equation}
Here, $\beta = 1/\mbox{k}_BT$, and $\left\{E_j^+({\bf k}), j=1,4\right\}$ are the positive eigenvalues of
 ${\bf A(k)}$. Let us put $t=1$. Now, all the energies (or parameters with units of energy) are in the units of $t$.
We find the self-consistent equations for $\Delta$ and $\Delta^\prime$ by minimizing $\Phi$ with respect to
$\Delta$ and $\Delta^\prime$. These are as follows.
\begin{eqnarray}
 \Delta & = & \frac{1}{2J}\frac{1}{4L}\sum_{\bf k}\sum_{j=1}^4 \frac{\partial E_j^+({\bf k})}{\partial\Delta}
           \mbox{tanh}\left(\frac{\beta E_j^+({\bf k})}{2}\right) \label{Delta_eq}\\
 \Delta^\prime & = & \frac{2}{J^\prime}\frac{1}{4L}\sum_{\bf k}\sum_{j=1}^4 \frac{\partial E_j^+({\bf k})}
                    {\partial\Delta^\prime}\mbox{tanh}\left(\frac{\beta E_j^+({\bf k})}{2}\right)
\label{Deltaprime_eq}
\end{eqnarray}
Since \( \partial\Phi/\partial\mu = -N\), where $N$ is the total number of electrons, we get the following
equation for the chemical potential.
\begin{equation}
 \delta = -\frac{1}{4L}\sum_{\bf k}\sum_{j=1}^4 \frac{\partial E_j^+({\bf k})}{\partial\mu}
           \mbox{tanh}\left(\frac{\beta E_j^+({\bf k})}{2}\right)
\label{mu_eq}
\end{equation}
The hole doping, $\delta = 1 - N/4L$.
Solving these sets of equations self-consistently gives us $\Delta$, $\Delta^\prime$ and $\mu$ as a function of
$\delta$, for given values of $\alpha$, $J$, $\beta$ and the phase angles $\theta_x$ etc.
\paragraph{The results of the mean field theory :}
We solve equations (\ref{Delta_eq}), (\ref{Deltaprime_eq}) and (\ref{mu_eq}) self-consistently for different values 
of $\delta$. We are interested in both the hole as well as electron doping for a given $\alpha$.
It is clear from the band structure that the hole doping for $\alpha$ is same as the electron doping for
$-\alpha$. Therefore, we have considered only the hole doping for both positive as well as negative $\alpha$.
In all our computations, we use $t=1$, $J=0.3$ and $|\alpha| =1.25$. The value of $J$ for SrCu$_2$(BO$_3$)$_2$
is roughly 70$^0$K. The ratio of $J$ to $t$ is tentative, and taken to be roughly same as that for the 
high-T$_C$ superconductors. Though we have four phases, $\theta_x$,
$\theta_y$, $\theta_{x+y}$ and $\theta_{x-y}$, only three relative phases are relevant. Therefore, we keep
$\theta_x = 0$. We have to find out those values of $\theta_y$ etc. for which the free energy is minimized,
and see how things evolve as a function of $\delta$.

\begin{figure}[h]
 \centerline{\epsfig{file=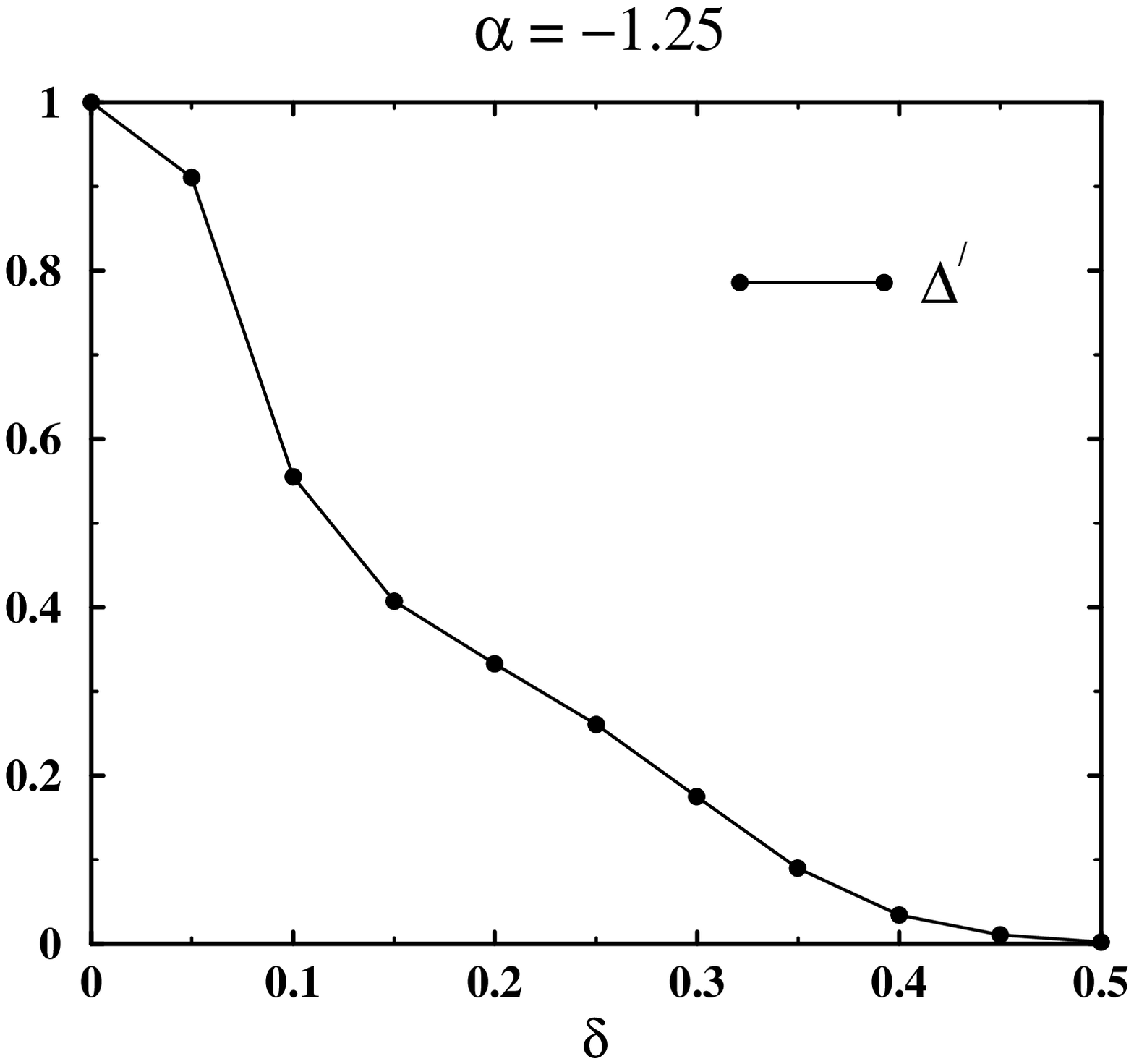,width=8.5cm}\hfill
             \epsfig{file=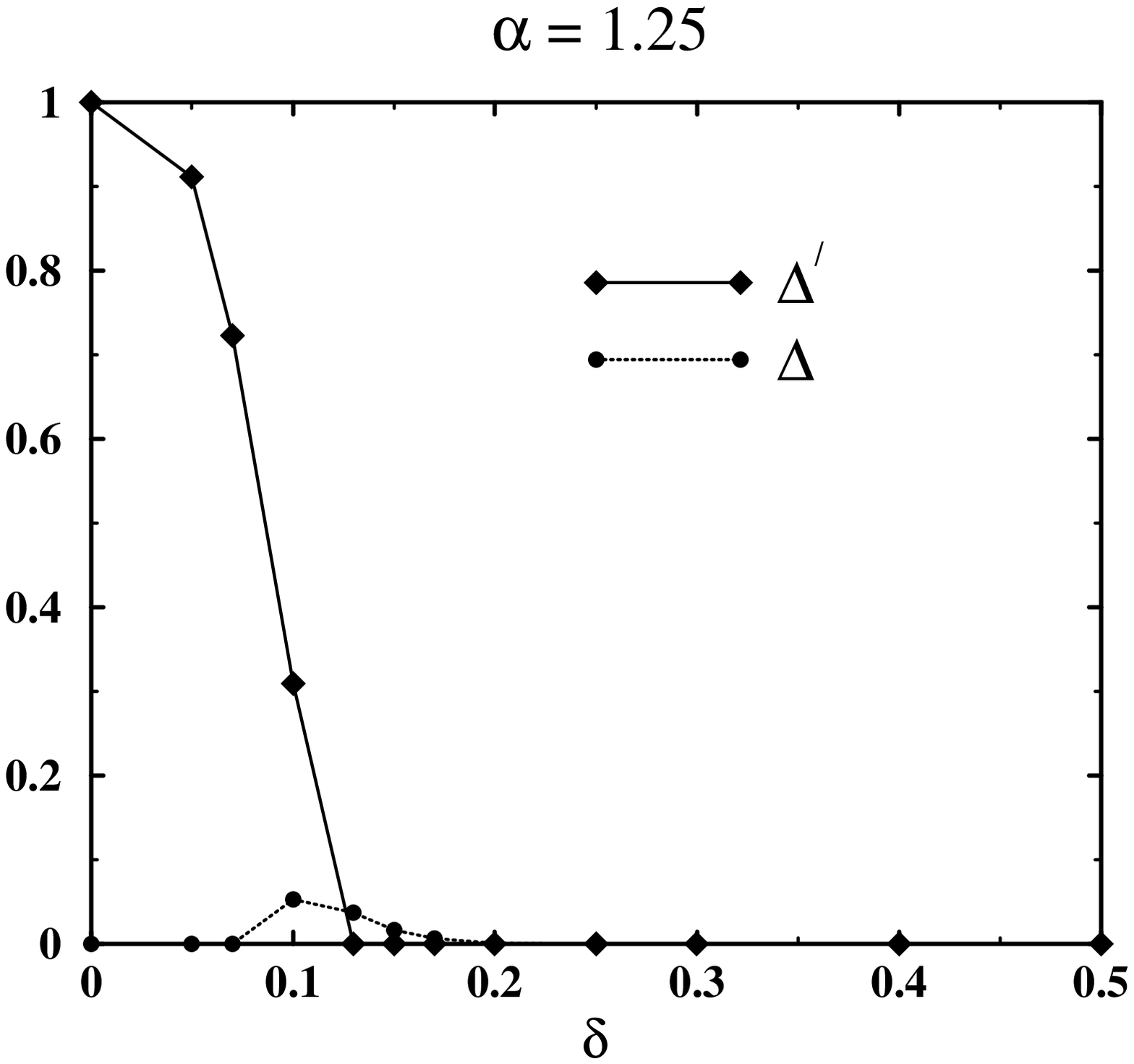,width=8.5cm}}
 \caption{The variation of $\Delta$ and $\Delta^\prime$ with doping, $\delta$, in the ground state. For
$\alpha=-1.25$, $\Delta$ is not shown, since it is identically zero.}
 \label{op_delta}
\end{figure}

Let us discuss the zero temperature ($\beta \rightarrow\infty$) case first. Fig.(\ref{op_delta}) shows the 
 variation of $\Delta$ and $\Delta^\prime$ with respect to $\delta$, for $\alpha = -1.25$ and 1.25 at zero
temperature.
For $\alpha = -1.25$, the minimum of free energy occurs for $\theta_{x+y} - \theta_{x-y} = \pi$ regardless of the
 values of $\theta_x$ and $\theta_y$, and $\Delta$ is identically zero. For $\alpha = 1.25$, the minimum of
free energy corresponds to $\theta_x = 0$, $\theta_y = \pi$, $\theta_{x+y} = 0$ and $\theta_{x-y} = \pi$.
It is a weak minimum as many other choices of the phases have similar values of the free energy. Nevertheless,
this choice of phases appears to be the minimum.
It is interesting to note that for $\delta = 0$,  the diagonal bond order parameter, $\Delta^\prime$, is 
one and $\Delta$ is zero (and is independent of the phases $\theta_x$, $\theta_y$, $\theta_{x+y}$ and $\theta_{x-y}$).
 {\em Thus, the RVB mean field theory at half filling
exactly reproduces the known dimer ground state of the SS model}.

To consider superconductivity in our mean field theory, we define a physical order parameter,
$\Delta_{\rm SC} = F_B\Delta_{\rm MF}$. Here, $\Delta_{\rm MF}$ is the mean-field order parameter
($\Delta$ or $\Delta^\prime$ whichever is larger for a given doping), and $F_B$ is a
bosonic mean field. Such an order parameter can be understood
in the framework of slave boson approach\cite{ubbens_lee},
 where the off-diagonal order parameter of the physical electrons is
described as $\left<b_ib_jf_{i\sigma}^\dagger f_{j{\sigma^\prime}}^\dagger\right>$. Here $b$'s are the slave
boson fields, and $f$'s are the fermionic objects. In a mean field decoupled theory, this is like
$\left<b_ib_j\right>\left<f_{i\sigma}^\dagger f_{j{\sigma^\prime}}^\dagger\right>\equiv F_B\Delta_{\rm MF}$. 
The bosonic order parameter, $F_B$, is a function of temperature and doping, and goes roughly like $\delta$. 
The superconducting 
transition temperature, T$_{\rm SC}$, is the temperature where either $\Delta_{\rm MF}$ or $F_B$ vanishes first while
increasing the temperature. For low doping, $\Delta_{\rm MF}$ is large, therefore, the T$_{\rm SC}$ is same as the bose
condensation temperature, T$_{\rm BC}$, for the bosonic field. Some estimates of T$_{\rm BC}$ have been made earlier 
while studying $t-J$ model in the context of the high-T$_C$ superconductivity\cite{ubbens_lee}. 
We roughly estimate it by considering an approximate dispersion of
the form, $k_x^2 + k_y^2 + \frac{1}{\gamma}k_z^2$, with the z-axis anisotropy $\gamma\sim 30$. We get
T$_{\rm BC} \approx 4\pi\rho^*\delta (1-\delta)/[2 + $Log$(4\gamma/\pi)]$. Here, $\rho^*$ is the density of states 
at the energy where two middle bands touch, from the side where dispersion is quadratic, and is a measure of
the curvature of the band. For $|\alpha| = 1.25$, $\rho^* \approx 0.1$, thus T$_{\rm BC} \approx 0.22\delta(1-\delta)$. 

One comment should be made regarding the interpretation of the result $\Delta=0$ and $\Delta^\prime\neq 0$ 
(for $\alpha < 0$). While at half filling this implied the dimerized insulating state, away from half filling it {\em
must} be interpreted as superconductivity. The BCS type wavefunction implies the fermion pairing in real space, \[
\left<c_{i\uparrow}^\dagger c_{j\downarrow}^\dagger\right> \sim \sum_{\bf k} \mbox{e}^{i{\bf k}\cdot 
(\mbox{\bf r}_i-\mbox{\bf r}_j)}\frac{\Delta_{\bf k}}{\sqrt{\Delta_{\bf k}^2 + (\epsilon_{\bf k} - \mu)^2}}\;\;, \] and it
extends over a range of lattice constants (due to the non-trivial {\bf k} dependence of $\epsilon_{\bf k}$ away from
$\delta = 0$), despite the mean field Hamiltonian having $n.n.$ pairing only. A similar remark holds for the four fermi
operator that determines the superconducting ODLRO of Yang, namely \( \left<c_{i\uparrow}^\dagger c_{j\downarrow}^\dagger
c_{j^\prime\downarrow}c_{i^\prime\uparrow}\right>\neq 0 \) for $|\mbox{\bf r}_i - \mbox{\bf r}_j| >> 1$.

\begin{figure}
 \centerline{\epsfig{file=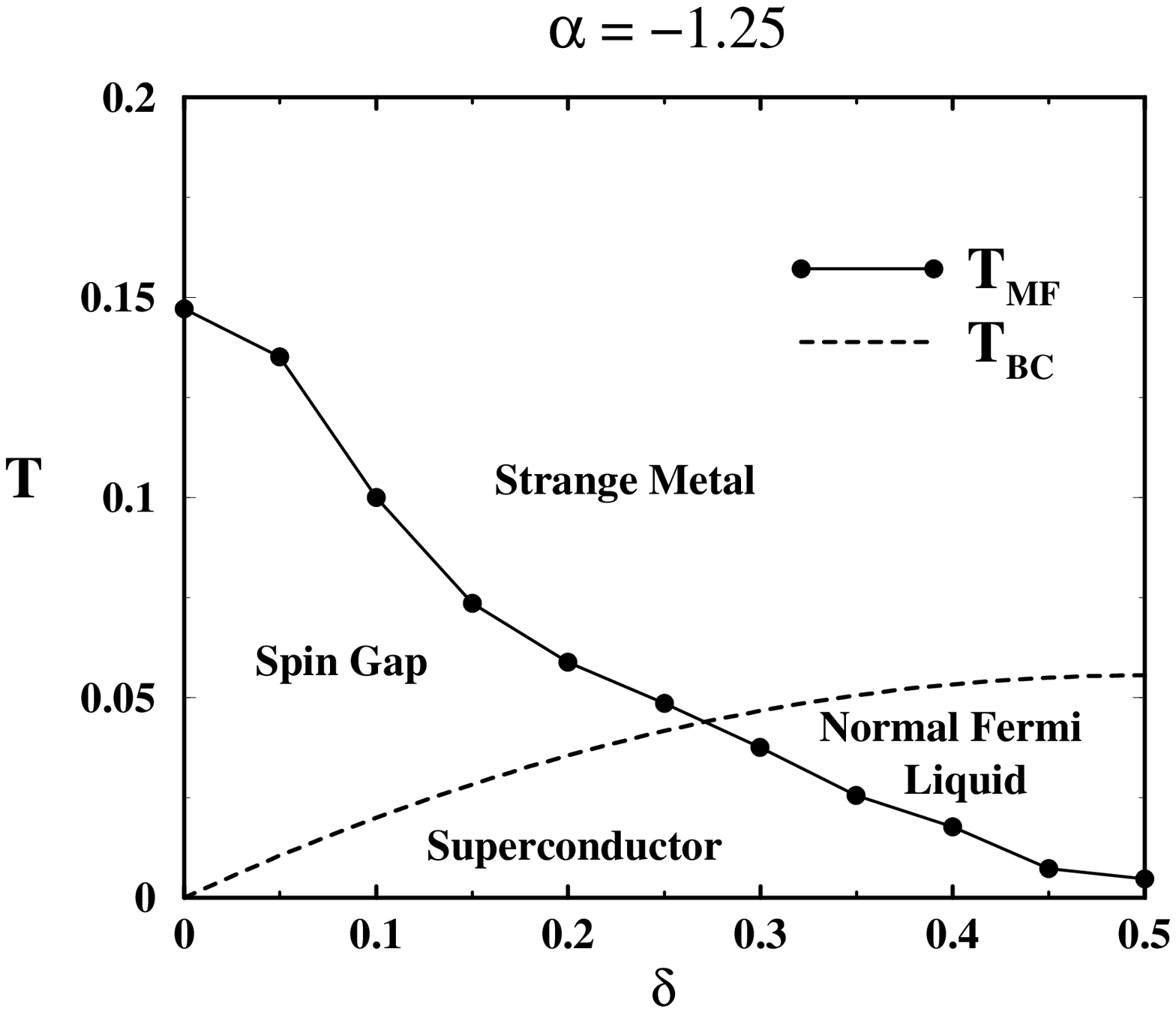,width=8.5cm}\hfill
             \epsfig{file=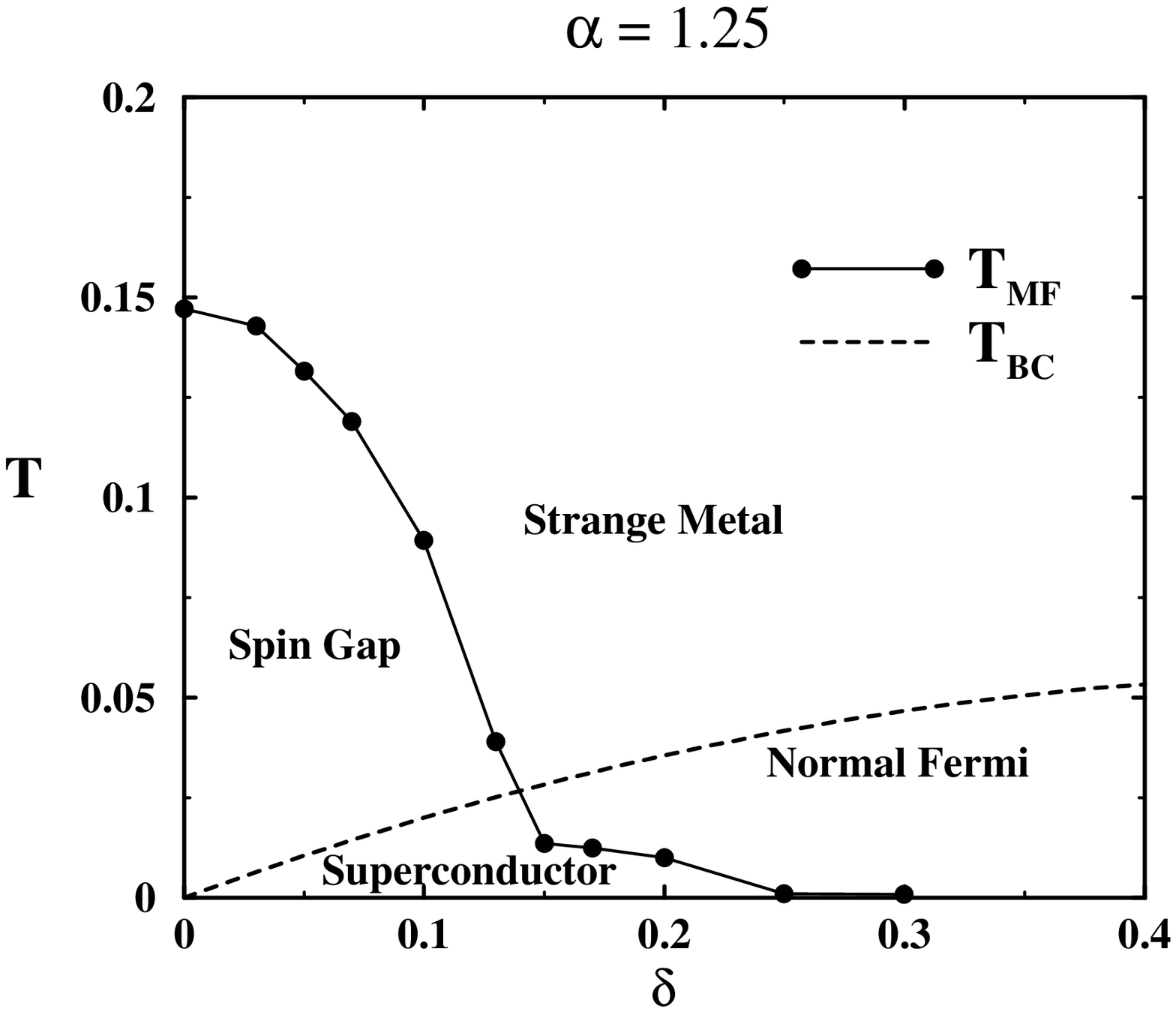,width=8.5cm}}
 \caption{The phase diagram for the negative as well positive values of $\alpha$. 
           The lines of the estimated Bose condensation 
          temperature, T$_{BC}$, and the computed mean field temperature, T$_{MF}$, divide the T-$\delta$ plane into
          four physically distinct regions. Each of these is appropriately labeled.} 
 \label{phase_diagram}
\end{figure}

Fig.(\ref{phase_diagram}) shows the phase diagram in the T-$\delta$ plane, as estimated from our RVB mean field
theory. The temperature, T$_{\rm MF}$, is where
$\Delta_{\rm MF}$ vanishes. The estimated T$_{\rm BC}$, and the computed T$_{\rm MF}$ are plotted as a function of $\delta$. The
common region under these two curves is the superconducting phase bounded by
critical lines. As usual all the remining lines should be viewed as
crossover lines rather than critical  lines. 
Among the remaining three regions of T-$\delta$ phase
diagram, the low doping region below T$_{\rm MF}$ and above T$_{\rm BC}$ is the spin gap phase with a suppressed density of states manifested in susceptibility as well as optical conductivity.
 Similarly, the high doping
region is the normal fermi liquid.  There is a region which is usually referred to as the strange metal
phase, as shown in the Fig.(\ref{phase_diagram}) with linear resistivity.
 Also, the phase diagram is similar for both positive as well
negative values of $\alpha$. From this phase diagram, we estimate the optimal value of the superconducting transition
temperature, T$_C\sim 10^0$K. 

In conclusion, the system considered here has a rather rich history. It may also have an important future 
since  under doping it might be  the much sought after low $T_c$ RVB superconductor, with  linear 
resistivity down to 10 K and other such exotic properties, rather than a conventional phononic BCS superconductor.

\section{Acknowledgements}
BSS would like to thank the   organizers of the  Nishinomiya Yukawa  Symposium for the warm hospitality. 
He thanks Professors E. Abrahams, P. Nozi\`{e}res, M. Schl\"{u}ter,  H. Kageyama,  K. Ueda  and J. Goplakrishnan for 
useful discussions   spread over various years.

{\it $^\dagger$Article based on an invited  talk given by BSS at the 16th Nishinomiya Yukawa memorial symposium
on `` Order and Disorder in Quantum Spin Systems'', Nov 13-14, 2001 at Nishinomiya}.

$^*$ The mean field theory of the doped dimer in Section 2. is joint work with B Kumar.

\begin{thebibliography}{**}

\bibitem{ss} B  S Shastry and B. Sutherland, Physica {\bf B 108}, 1069 (1981)
\bibitem{mef} M E Fisher, in Summary talk at the 1992 USA  Japan  meeting at the Institute of Theoretical Physics,
Santa Barbara. Professor Fisher emphasized here and elsewhere, the crucial role played by the two dimensional Ising model solution
of Onsager in the development of the theory of critical phenomena.
 Our colleague at IISc, Professor Chandan Dasgupta, has remarked in a similar vein
that,  if not for  Baxters solution of the two dimensional q state Potts model order parameter, a truly subtle object, we might even today be debating the order of the phase transition in the system.
\bibitem{ss2} B Sutherland and B S Shastry,  J. Stat. Phys. {\bf  33}, 477 (1983).
\bibitem{mu} S. Miyahara and K. Ueda, Phys. Rev. Lett {\bf 82}, 3701 (1999), (2000) cond-mat /0004260.
\bibitem{knetter} C Knetter, A B\"{u}hler, E M\"{u}ller-Hartmann and G Uhrig, Phys Rev Lett {\bf 85} 3858 ( 2000).
\bibitem{neutron} H Kageyama, M Nishi et al, Phys Rev Lett {\bf 84} 5876 (2000)
\bibitem{nmr} H Kageyama, K Yoshimura et al, Phys Rev Lett {\bf 82}3168 ( 1999).
\bibitem{esr} H Nojiri et al, J Phys Soc Japan {\bf 68} 2906 ( 1999).
\bibitem{raman} P Lemmens et al, Phys Rev Lett {\bf 85} 2605 (2000).
\bibitem{thermal} M Hoffmann et al, Phys Rev Lett {\bf 87} 047202 ( 2001).
\bibitem{excitations} K Totsuka, S Miyahara and K Ueda, Phys Rev Lett {\bf 86} 520 ( 2001), Y Fukumoto, 
cond-mat/000411 (2000).
\bibitem{plateauexp} K Onizuka et al, J Phys Soc Japan {\bf 69}, 1016 ( 2000).
\bibitem{plateaus1}Y Fukumoto, cond-mat/0012396, T Momoi and K Totsuka, Phys Rev {\bf 62} 15067 ( 2000),
\bibitem{plateaus2} G Misguich, Th Jolicoeur and S  Girvin, Phys Rev Lett {\bf 87} 097203 ( 2001).
\bibitem{kk} A  Koga and N Kawakami ,  Phys Rev Lett {\bf 84} 4461 ( 2000), cond-mat/0003435 (2000).
\bibitem{oitmaa}Z Weihong, J Oitmaa and C Hamer, cond-mat/0107019 ( 2001).
\bibitem{uwe}U L\"{o}w and E  M\"{u}ller-Hartmann, cond-mat/0104385 ( 2001).
\bibitem{cms} C Chung, J Marston and S Sachdev, Phys Rev {\bf 64} 134407 ( 2001).
\bibitem{carpentier} D Carpentier and L Balents, cond-mat/0102218 ( 2001).
\bibitem{am} M Albrecht and F Mila, Europhys Lett {\bf 34} 145 ( 1996).
\bibitem{theoretical} N Surendran and R Shankar, cond-mat/0112507 ( 2001),  
\bibitem{mh}E M\"{u}ller-Hartmann, R R P Singh, C Knetter and G Uhrig, Phys Rev Lett {\bf 84} 1808 ( 2000). 
\bibitem{nozieres} We owe this comment to Professor P N\`{o}zieres.
\bibitem{mgb2}J Nagamatsu,  N Nakagawa, T Muranaka, YZenitani and J Akimitsu, Nature, {\bf 410} 63, (2001), 
J Kortus, I I Mazin, K D Belashchenko, V P Antropov and L L  Boyer,cond-mat/011446
\bibitem{BZA} G. Baskaran, Z. Zou and P. W. Anderson, Solid State Commun. {\bf 63}, 973 (1987)
\bibitem{kotliar_liu} G. Kotliar and J. Liu, Phys. Rev. B {\bf 38}, 5142 (1988)
\bibitem{ubbens_lee} M. U. Ubbens and P. A. Lee, Phys. Rev. B {\bf 49}, 6853 (1994)

\end{thebibliography}
\end{document}